# Removal of clouds from satellite images using time compositing techniques

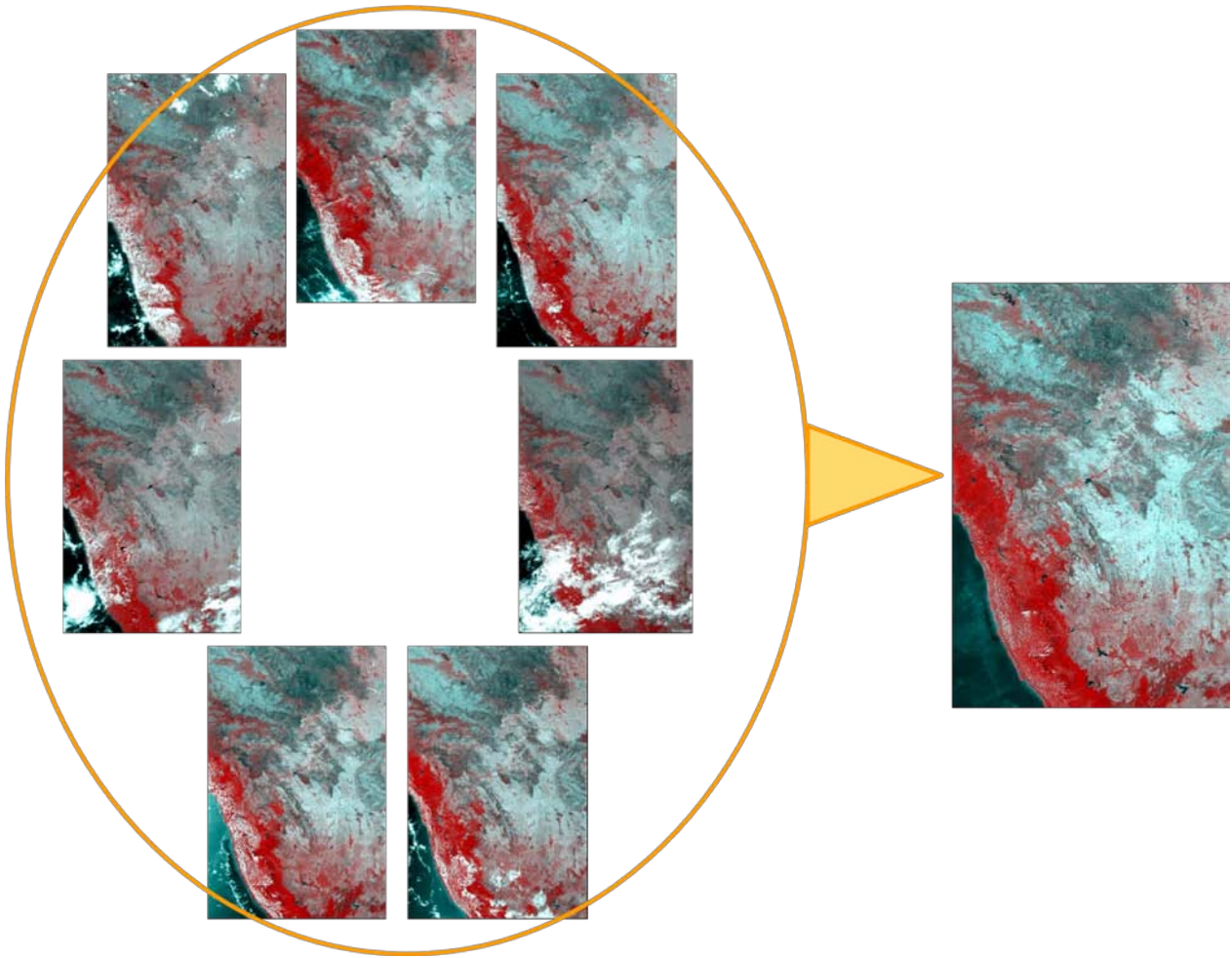

Project work by,

**\* M Atma Bharathi, TR Nagashree, P Manavalan, PG Diwakar,**

Regional Remote Sensing Service Centre - Bangalore, ISRO.

\* Corresponding author: atmabharathi@isro.gov.in

# Abstract

Clouds in satellite images are a deterrent to qualitative and quantitative study. Time compositing methods compare a series of co-registered images and retrieve only those pixels that have comparatively lesser cloud cover for the resultant image. Two different approaches of time compositing were tested. The first method recoded the clouds to value 0 on all the constituent images and ran a 'max' function. The second method directly ran a 'min' function (without recoding) on all the images for the resultant image. The 'max' function gave a highly mottled image while the 'min' function gave a superior quality image with smoother texture. Persistent clouds on all constituent images were retained in both methods, but they were readily identifiable and easily extractable in the 'max' function image as they were recoded to 0, while that in the 'min' function appeared with varying DN values. Hence a hybrid technique was created which recodes the clouds to value 255 and runs a 'min' function. This method preserved the quality of the 'min' function and the advantage of retrieving clouds as in the 'max' function image. The models were created using Erdas Imagine Modeler 9.1 and MODIS 250 m resolution images of coastal Karnataka in the months of May, June 2008 were used. A detailed investigation on the different methods is described and scope for automating different techniques is discussed.

# Introduction

With time, the number of remote sensing satellites and their resolutions have increased steadily. Sensors with wide swath have refined the temporal resolutions to less than 5 days. Yet collecting images devoid of cloud cover is at the mercy of nature. Opaque clouds hamper the usability of these images. Indices like NDVI have been used to characterize the extent of vegetation on a particular scene. Since the NDVI is derived from satellite optical remote sensing data, cloud contamination not only significantly influences the precision of its application but also makes it unfeasible to obtain completely cloud-free NDVI images for a limited period over a large area or in mountainous regions [1]. Hence there is a compelling requirement to develop techniques which can either remove clouds or at the least mask them from the study area.

## Objective

The project required to develop weekly maps of vegetation extent and density in the coastal region of Karnataka state, India. The study area being a tropic and a coastal zone, it had high cloud cover during the monsoon period. Hence a technique to remove clouds without disrupting the information content regarding extent vegetation was developed.

## Materials used

MODIS 250m channels in bands 7 (2105 – 2155 nm), band 2 (841 – 876 nm), band 1 (620 – 670 nm) were used. False colour composites were produced with band combination 2,1,1. Bands 2 (NIR) and 1 (Red) were used to create NDVI images. The images were co-registered and Erdas Imagine Modeler 9.1 was used to create different time compositing models.

# Methodology

Time compositing methods function by comparing corresponding pixels in a finite set of co-registered images. Based on the model used, different parameters are compared pixel wise and the best fit is copied to the resultant image.

Figure 1 is an initial flow chart depicting the steps followed. Two disparate methods were tested. The boxes in the left belong to the first method and that in the right belong to the second method. At the end, a hybrid method was evolved retaining the advantages of both methods. In the following pages these methods and the results are explained in detail.

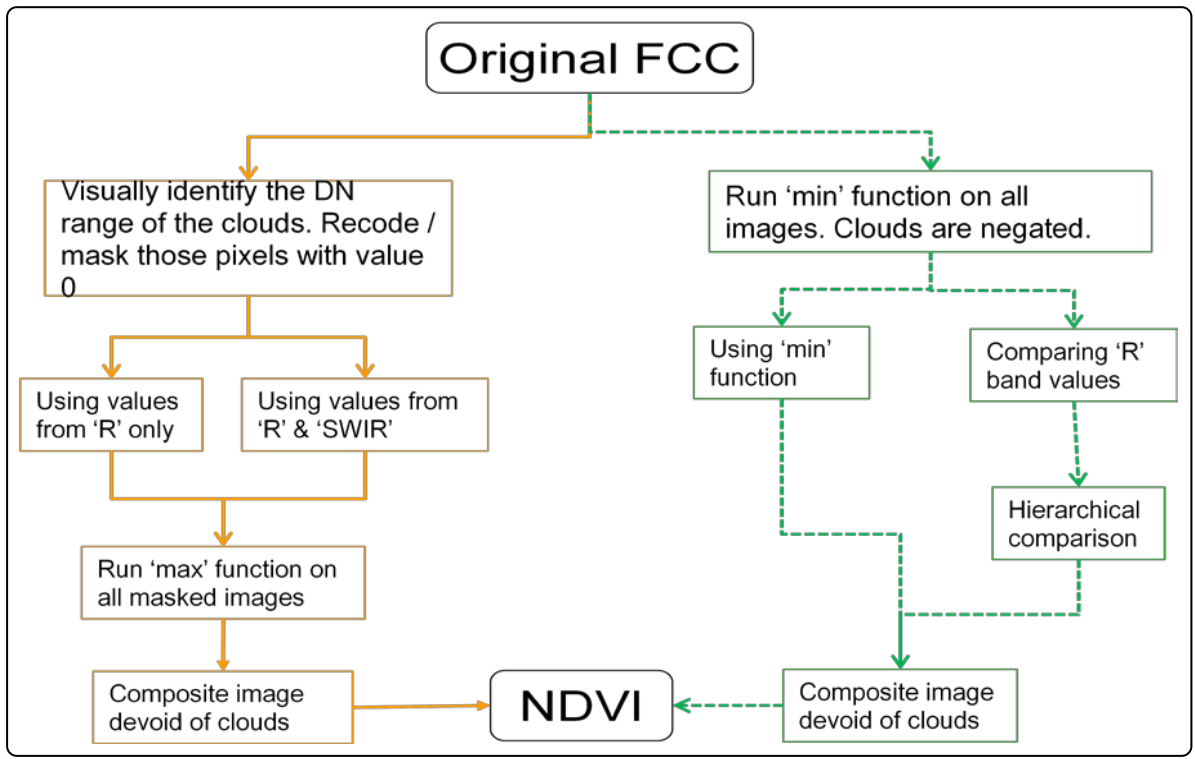

**Figure 1: Initial flow chart displaying the steps involved in the two different approaches to cloud removal**

## Method 1: Cloud removal by masking clouded pixels.

In this method, pixels with clouds are identified and recoded to value 0 on all the constituent images. By selecting only that pixel with maximum DN aggregate out of the corresponding pixels in the image set, we can negate the masked pixel and hence the cloud. This pixel is then copied in the new composite image. By applying this function to the entire image, all the masks can be negated and a final composite image devoid of cloud cover can be generated.

### Steps

Of the 3 bands, SWIR, NIR & R, red has the least wavelength and hence greater scattering making it to capture all the clouds. It was found that clouds could occur within the bracket [150 to 255] DN in red band. An algorithm to search for those pixels with DN greater than 150 in red band and to recode them to DN 0 on all bands was created. When an image was fed, the output was a cloud masked image. In the next step, a maximum function was run on these cloud masked images to obtain a composite image devoid of clouds.

### Observation

With using the 'R' band, clouds were masked efficiently. In addition a few pixels from the fallow land too were masked since their DN in Red band overlapped the cloud bracket. Since SWIR has longer wavelength, it would skip those clouds which can scatter the Red band and reach the ground below. Thus for pixels with clouds, the DN in SWIR would be lesser than Red, while in fallow land, the SWIR DN would be higher than Red. This condition was used and the algorithm was modified suitably to screen out those pixels whose DN in Red was lesser than that of SWIR.

### The Model

A model was built in Erdas Imagine Modeller 9.1 which applies the refined cloud mask on 7 constituent images; runs a maximum function on the cloud masked and extracts an NDVI image from the cloud free composite image. Figure 2 displays the model.

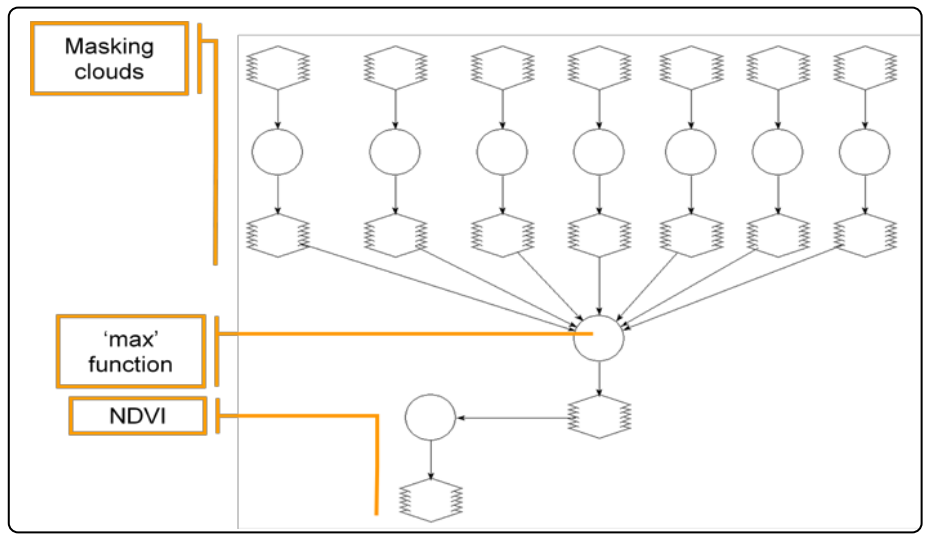

**Figure 2: Model to cloud mask images, create a composite image using maximum function and to extract an NDVI image.**

The resultant NDVI image was sliced and the ranges were colour coded. Table 1 displays the ranges and the colours assigned.

| Range | Class | Colour |
|-------|-------|--------|
| -1 to 0 | Water | Blue |
| 0 to 0.09 | Clouds | Grey |
| 0.09 to 0.25 | Fallow land | Pale yellow |
| 0.25 to 0.5 | Moderate vegetation | Mild green |
| 0.5 to 1 | Dense vegetation | Dark green. |

**Table 1: Table showing NDVI slices and the colour coding.**

## Method 2: Cloud removal using minimum function

Of all the different features present in the image, clouds have maximum DN in all 3 bands. By copying only that pixel which has minimum DN aggregate from the image set, we get a pixel which represents a feature that would not be a cloud. A model was developed to house the minimum function and was run on all the constituent 7 images.

### Observation

The minimum function should ideally sum the DN values all the bands for a pixel, compare this sum with the rest of the pixel sums to determine the minimum pixel. But it was observed that it rather compared each band's DN individually and copied the least value to the destination pixel. Table 2 shows a table with the constituent images and the destination pixel when subjected to minimum function.

| Image | SWIR (B1) | NIR (B2) | R (B3) |
|-------|-----------|----------|--------|
| 2$^{nd}$ may | 81 | 181 | 68 |
| 3$^{rd}$ may | **44** | **158** | 53 |
| 4$^{th}$ may | 76 | 175 | **49** |
| Resultant image | 44 | 158 | 49 |

**Table 2: Table showing the synthetic pixel created by the minimum function**

This is a serious error since the purity of a pixel and hence the entire image is lost. The resultant is rather a synthesized image than a composite image. In addition, as vegetation progresses in growth, the reflectance in NIR band increases considerably and that in Red band decreases mildly. Thus selecting the pixel with least reflectance aggregate would result in a pixel with least vegetation. Thus for a project to map the maximum extent of vegetation in a time frame, the algorithm should compare only the Red band values and copy the entire pixel with least reflectance in red band.

## The Model

To incorporate the refinement, the constituent images had to be compared two at a time hierarchically and a model was developed to automate this. At the end, the model extracted an NDVI image. The same class intervals and shades were used to slice the image and to colour code it. Figure 3 displays the consolidated model for refined minimum function method.

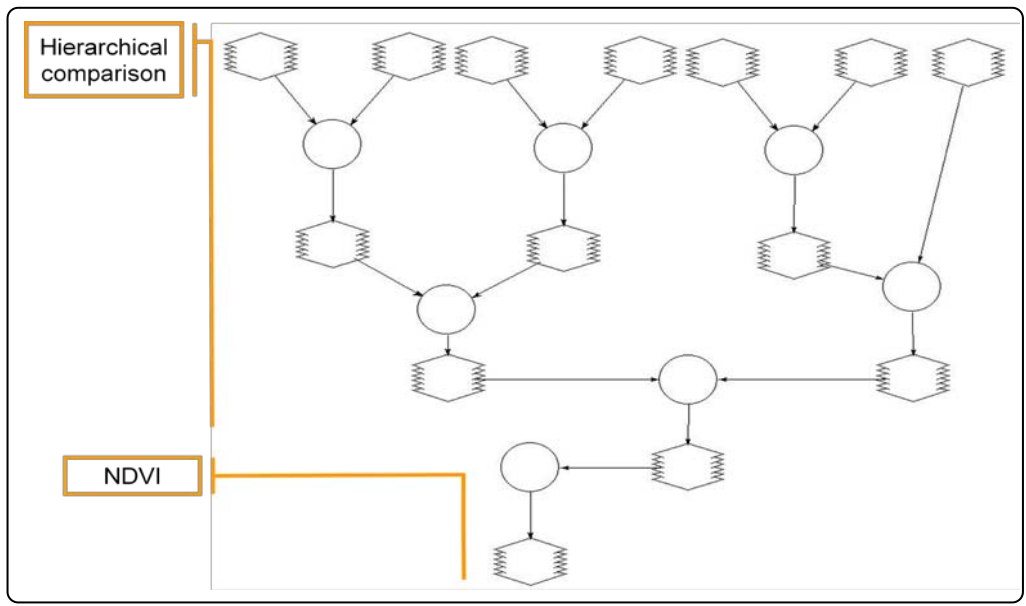

**Figure 3: Consolidated model for 7 input images for the refined minimum function method.**

# Results

Both the models were run on several sets of 7 images of consecutive dates. The images were analysed and compared and the results are presented below.

## Analysis of maximum function method

The resultant image was fairly devoid of clouds but was highly coarse in texture. The cloud masking algorithm could not mask the peripheral pixels around the clouds. These pixels with brighter DN were retained in the resultant image causing the coarseness in texture.

Figure 4 (colour plate) shows one of the constituent images, the resultant image when the model was run on a set of 7 images covering the dates May 15[th] to 21[st] 2008 and the corresponding NDVI image. The white slanted line is a carryover of a missing path in one of the constituent images. The southwest portion along the coast is covered with dense vegetation but the composite has produced a highly mottled texture. The NDVI extracted from that image is displayed adjacent. The majority of the study area is classified as fallow land, displayed in pale yellow. Moderate and dense vegetation is displayed in mild and dark shades of green. Water bodies are in blue and remnant clouds in black.

## Analysis of refined minimum function method

The resultant image was devoid of clouds when applied during the month of May. The texture of the image was comparatively smoother, yet the coastal vegetation appeared mottled. The minimum function inherently favours darker pixels. While the clouds are removed, their shadows are preserved in the resultant image. When the constituent images are not accurately co-registered this problem gets vicious. The mild displacements of dark features get exaggerated. Figure 5 (colour plate) displays a water body in one of the original images and the corresponding water body in the composite image. A growth in the size can be seen as the tone of the feature is dark. Further the coastline gets modified. As the minimum function favours the darker pixels of the sea, the resultant image shows the coast retreated inland. Figure 6 (colour plate) displays a composite image and one of the source images overlaid. The source image is displayed from right up to the green vertical line and the composite image is displayed from the green line to the left end of the image. The yellow line shows the actual direction of coast, while the retreated cost can be seen in the composite image.

The composite image for the same dates from May 15th to 21st and the corresponding NDVI extract is displayed in figure 7 (colour plate).

## Comparison of the two methods

Comparing the outputs in figure 4 and 7 depicts the minimum function method as better. The models were run on other sets of images during the month of May and consistently, minimum function gave better results. The NDVI image from the minimum function composite reflects the reality on ground closely. Figure 8 displays the histograms of NDVI images from the two methods. A large amount of variation in the number of pixels falling under each slice can be seen.

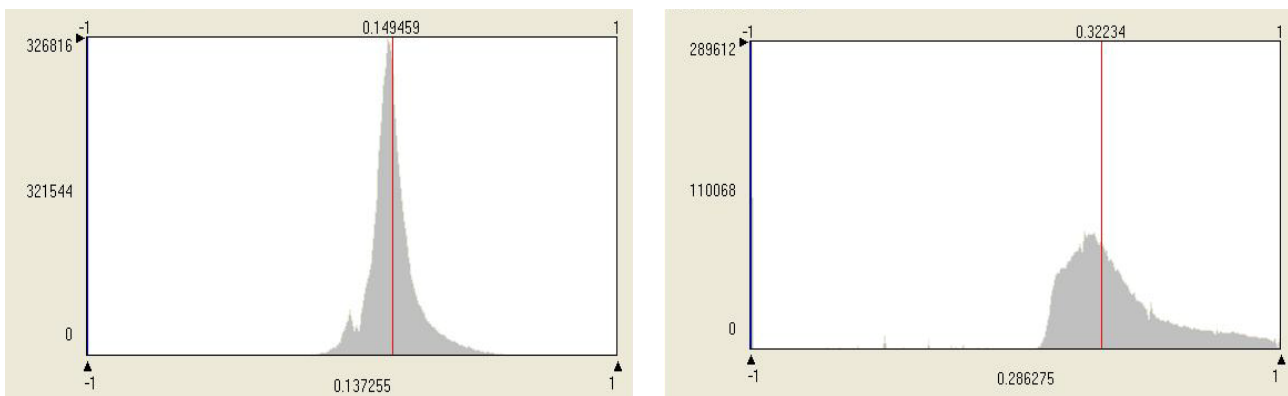

**Figure 8: Comparing the histograms of NDVI images created using the maximum function (left) and the minimum function models.**

It can be observed that the minimum function is more suitable for this study as it inherently favours the vegetation pixels. Further, this method does not require checking the range between which the clouds lie as opposed to the cloud masking method.

When the models were applied to images collected during the month of June, the south-west monsoon period, the resultant images contained fairly large amount of clouds. These clouds were persistent and static over the same pixels on all the constituent images. In the cloud masking composite, these pixels had a standard value 0 and retrieving them was simple. While in the refined minimum function composite, clouds varied over a range of DN values hindering their easy retrieval and slicing of extracted NDVI image. This problem limits the use of the method to the off-monsoon months.

# Evolution of a hybrid cloud removal method

The limitation of the refined minimum function method can be reduced by retaining the refined minimum function, yet masking the clouds in the images. In order to apply the function, the clouds were masked to the highest DN value 255 in all bands (8 bit quantization).

## Steps

The steps were similar to the first method. The DN bracket for the clouds were identified and recoded to value 255 in all bands. Next, the image pairs were compared hierarchically for the minimum DN in Red band and a composite image was created.

## Observation

The hybrid model was run on several image sets and found to give better results consistently. Under moderately clouded dates, the composite and NDVI extract were more accurate than that from the refined minimum method. With the hybrid method, slicing of NDVI was accurate since all negative values corresponded to water bodies, clouds were exact 0 and the positive values were sliced as in other methods.

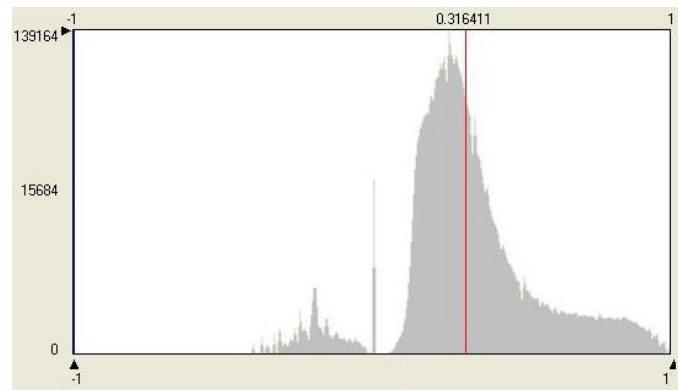

**Figure 9: Histogram of NDVI image from the hybrid method**

In figure 9, the sharp vertical bar corresponds to NDVI value 0 which are the remnant clouds in the composite image. The little mound with values less than 0 map the water bodies. The mean of the vegetated portion is at 0.316, falling under the moderate vegetation class.

# Discussion

The study was conducted based on an assumption that clouds are dynamic and would not be static over the same region throughout a time limit (here 7 days). Thus during the off-monsoon months, the composite was considerably devoid of clouds. But during the monsoon period, these clouds were found to accrete drastically masking out much of the study area. The cloud bodies with peak reflectance were found to be more persistent and were eventually carried over to the composite. Figure 10 (colour plate) shows the composite images from different methods for the period of June 8th to 14th 2008 and one of the source images. Clouds masked to 255 in the hybrid output can be negated from the NDVI outputs. One way to handle clouds during the monsoon period would be to increase the number of images from different dates inside the finite set. Using the composites as inputs for the same model would again yield a composite with much lesser cloud cover.

The hybrid method inherits the merits of both the cloud masking and refined minimum function methods. Throughout the study, the bracket [150 – 155] was found suitable for masking out the clouds. Yet, this can be refined by observing the first point of inflection occurring after the DN value 150 in the histogram of red band. During the monsoon period, the numbers of pixels with cloud cover increase dramatically and this can be seen as a sharp rise in the histogram of red band after value 150 as shown in Figure 11. By breaking the shape of histogram as a series of line segments, one could analyse the slope of those segments occurring after 150. The first segment which records a slope higher than a preset threshold (say >1), indicates the point of inflection. The DN value corresponding to this point can be used as the refined lower threshold for masking the clouds.

However, during the off-monsoon periods, the cloud cover was much lesser and no inflection could rather be seen in the histograms. Under this condition, the general bracket can be used to mask clouds.

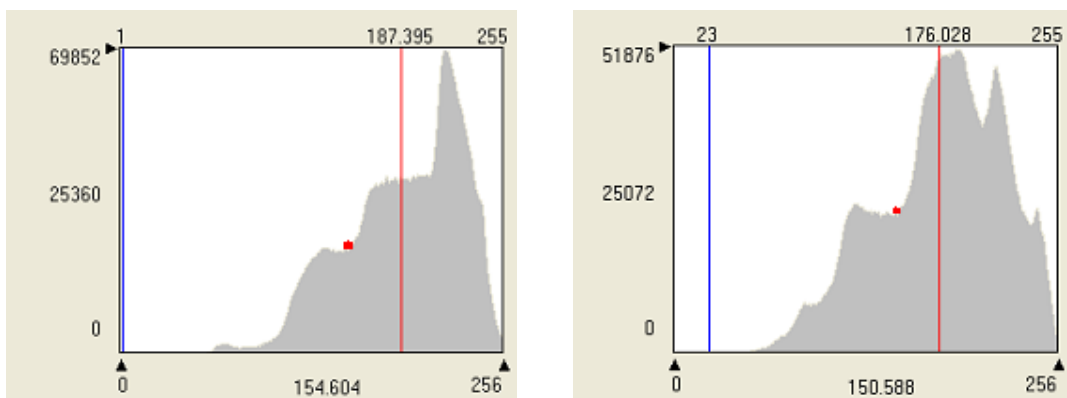

**Figure 11: Histogram of red band for images with profuse cloud cover. The dot on the graph shows the first point of inflection occuring after the DN value 150**

         This algorithm does not remove the shadows of the clouds. Yet due to the coarse resolution (250m) of the MODIS bands, shadows were rather dispersed and did not pose a problem in the composite image. Resampling the composite image to cubic convolution was found to give smoother and better results.

# Conclusion

The simple cloud masking method although theoretically perfect, fails miserably due to the 'max' function. Further, the max function would inevitably reduce the vegetative land cover. The refined minimum method gives results much better than the 'max' function method yet, becomes unsuitable when used during monsoon periods. The hybrid method remains suitable to be used under most conditions. Yet under blinding cloud cover (as during months of June), the hybrid method requires a larger set of data to produce a usable composite image.

         The hybrid method can be completely automated where the user could feed the finite set of co-registered images and the model would compute the composite. The cloud masking bracket can be refined using the point of inflection in the histogram.

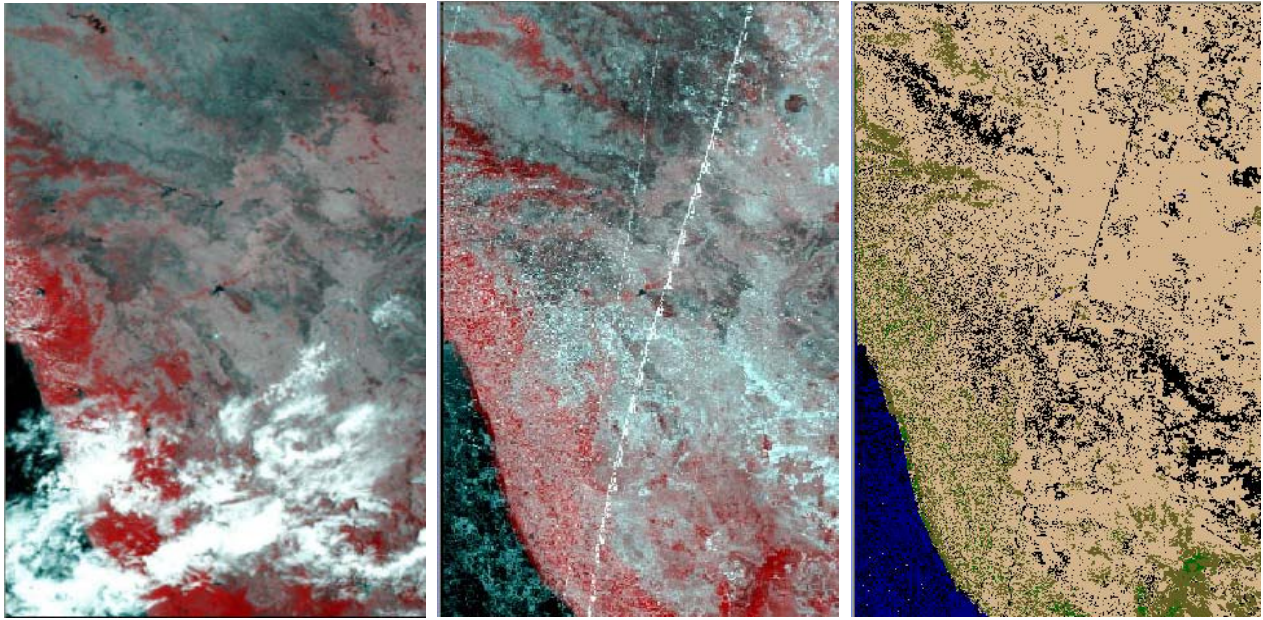

**Figure 4: Resultant image and NDVI extract using maximum function model for a set of 7 images covering May 15th to May 21st 2008**

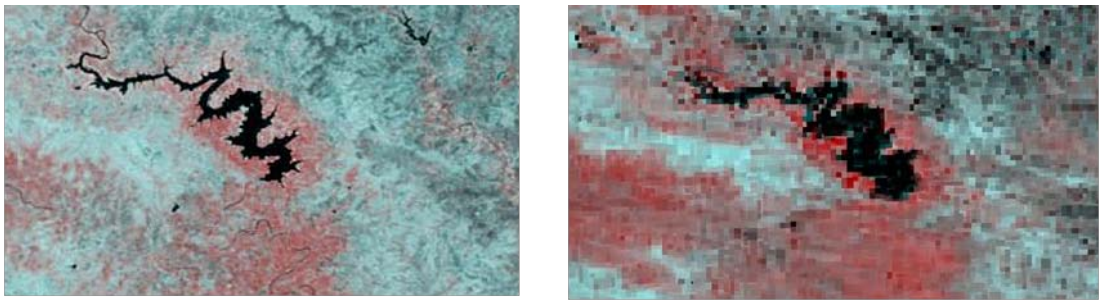

**Figure 5: The water body in the original image and the growth observed in the composite image**

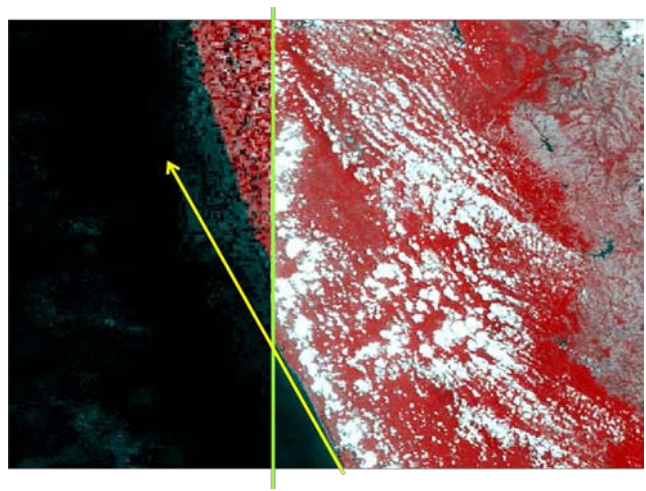

**Figure 6: The composite and original image overlaid. The retreat in coast line is seen and represented by the gap between yellow line and the land mass in the image.**

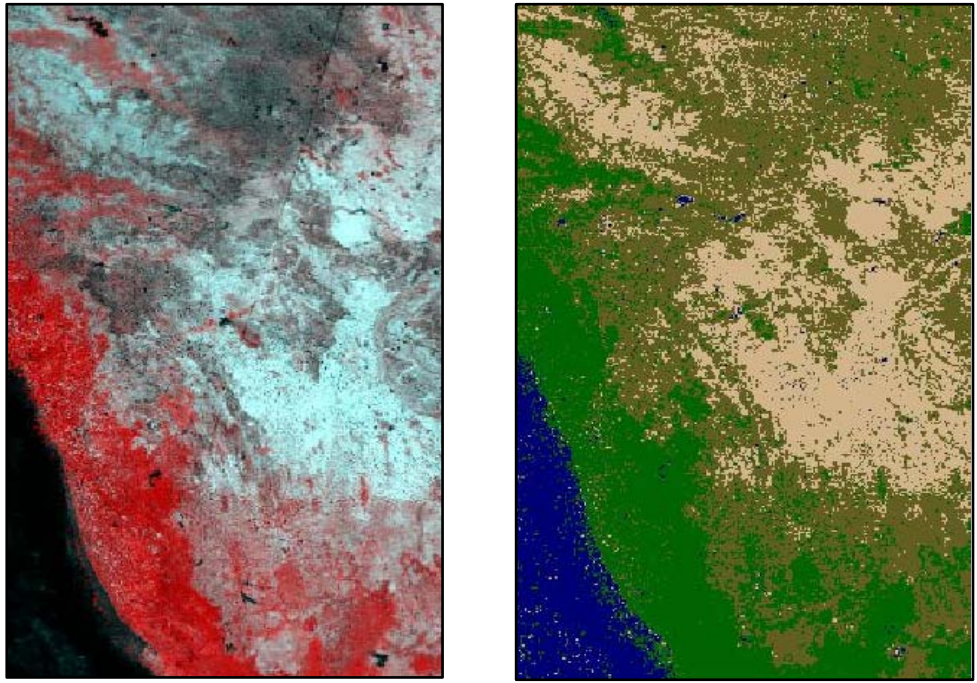

**Figure 7: Composite image from  refined minimum function method and the corresponding sliced NDVI image**

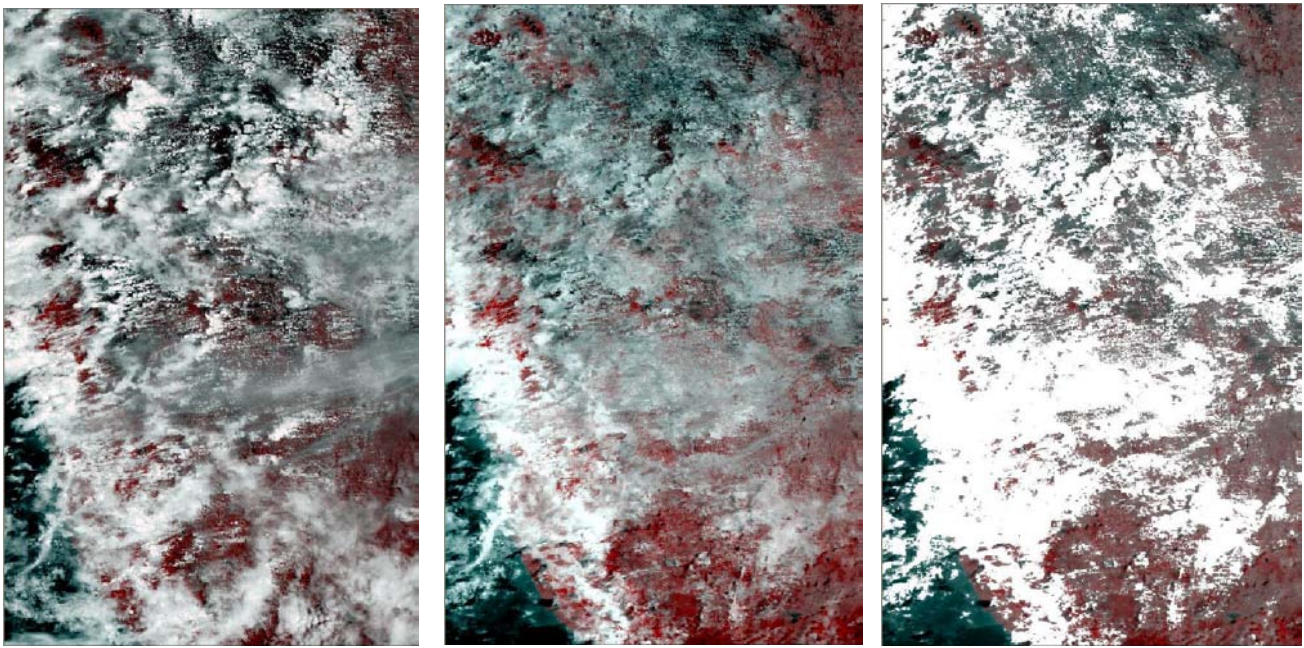

**Figure 10: Highly clouded dates of June, the composite from refined minimum function method, composite from the hybrid method.**